\documentstyle[aps,epsfig]{revtex}
\newcommand{\om}{\Omega_{\rm M}}
\newcommand{\ola}{\Omega_\Lambda}
\newcommand{\beq}{\begin{equation}}
\newcommand{\eeq}{\end{equation}}
\newcommand{\beqa}{\begin{eqnarray}}
\newcommand{\eeqa}{\end{eqnarray}}

\begin{document}
\twocolumn

\title{Determining the fraction of compact objects in the Universe using supernova 
observations}
\author{E. M\"ortsell\thanks{E-mail address: edvard@physto.se}, 
A. Goobar\thanks{E-mail address: ariel@physto.se}, 
and L. Bergstr\"om\thanks{E-mail address: lbe@physto.se}}
\address{Department of Physics, Stockholm University, \\
         Box 6730, S--113 85 Stockholm, Sweden}
\maketitle

\begin{abstract}
We investigate the possibility to determine the fraction of compact objects in 
the Universe by studying gravitational lensing effects on Type Ia supernova observations.
Using simulated data sets from one year of operation 
of the proposed dedicated supernova detection satellite 
SNAP, we find that it should be possible to determine the fraction of 
compact objects
to an accuracy of $\lesssim 5 \%$. 
\end{abstract}

\section{Introduction}\label{sec:intro}
Recent measurements of anisotropies in the cosmic microwave background radiation (CMBR) 
show that the Universe is very close to flat, i.e.,
$\Omega_{\rm tot}\approx 1$ \cite{boomerang,maxima}. Since observations of Type Ia 
supernovae (SNe) indicate that the expansion rate of the Universe is accelerating, the major
part of this total energy should have negative pressure, e.g., in the form of
the cosmological constant corresponding to
 $\ola\sim 0.7$ in a flat universe \cite{perlmutter,riess}, in agreement
with constraints on the matter density $\om$
from cluster abundances \cite{bahcall,carlberg} and large-scale structure \cite{2DF}. 
Thus, a  a concordance model with $\om \approx 0.3$ and $\ola \approx 0.7$ has emerged. 

The constitution of the total matter density is a matter of intense theoretical and 
experimental research. 
The energy density in baryonic matter as derived from Big Bang nucleosynthesis (BBN)
is given by $\Omega_{\rm b}h^2=0.019\pm0.0024$ \cite{burles} whereas 
CMBR measurements yield a somewhat higher value; $0.022<\Omega_{\rm b}h^2<0.040$
\cite{tegmark}.
Either way, the matter density in baryons is almost one order of 
magnitude smaller than the 
non-baryonic dark matter (DM) component. However, the BBN range for the baryon
density still means that most of the baryonic matter is also dark. 

There are
various possibilities for the dark baryons to hide, examples are warm 
gas in groups and clusters 
(which is difficult to detect at present \cite{gas}), 
or in the 
form of massive compact halo objects (MACHOs), where indeed there have been
detections \cite{macho,eros}. The long lines of sight to distant supernovae
means that they are  well suited to probe the matter content along the
paths of the light rays, where gravitational lensing may occur where
there are matter accumulations. Compact objects give more distinct
lensing effects,
enabling a distinction between diffuse matter and compact bodies along the light path.
In particular, it may be possible to
investigate  whether the halo fraction deduced for the Milky Way from 
microlensing along the line of sight to the
Large Magellanic Cloud, of the order of 20 \% \cite{macho}, 
is a universal number or if the
average cosmological fraction is larger or smaller.

Regardless of its constitution, we can classify dark matter according to its clustering 
properties. In this paper we will use the terminology of {\em smooth DM} for DM candidates
which tend to be smooth on subgalaxy scales, e.g., 
weakly interacting massive particles (WIMPs) such as neutralinos.
The term {\em compact DM} will be reserved for
MACHOs such as brown or white dwarfs and primordial 
black holes (PBHs) \cite{jedamzik}.

An advantage with gravitational lensing is that its 
effects can determine the distribution of dark matter independent of its 
constitution or its dynamical state.
The topic of  this paper concerns 
the use of the gravitational magnification of standard candles,
such as Type Ia SNe, 
to determine the fraction of compact objects in a cosmological context.

In an early study, Rauch \cite{rauch} concluded that with a sample of 1000 Type Ia SNe at 
redshift $z\approx 1$, one should be able to discriminate between the extreme cases of
all DM as smooth or in compact objects. More recent work (Metcalf and Silk \cite{metcalf},
Goliath and M\"ortsell \cite{goliath}) has shown that a sample of 50--100 SNe should be 
sufficient to make the same discrimination. Seljak and Holz \cite{seljak} have found 
that it should be possible to actually determine the fraction of compact objects to 20 \%
accuracy with 100--400 Type Ia SNe at $z=1$.
In this work we extend these studies by exploring the possibility to determine the fraction 
of compact objects using a future sample of Type Ia SNe distributed over a 
broad redshift range, with the intrinsic spread of absolute luminosity of the
SNe and expected measurement error for a proposed space-borne mission,
SNAP \cite{snapprop} taken into account.

In Sec.\,~\ref{sec:compact} we clarify the distinction between
compact and non-compact dark matter objects. In Sec.~\ref{sec:sn} we discuss gravitational lensing of SNe and in Sec.~\ref{sec:snoc},
we present the Monte-Carlo simulation package used to predict this effect. 
Sec.~\ref{sec:snap} is concerned with the proposed satellite telescope which will be able 
to produce the data sets used in this study and in Sec.~\ref{sec:res} we present our results. 
The paper is concluded with a summary in Sec.~\ref{sec:disc}.

\section{Compact Objects}\label{sec:compact}
For an object to be compact in a lensing context, we demand that it is contained within 
its own Einstein radius, $r_E$.
For a DM clump at $z=0.5$ and a source at $z=1$, $r_E\sim 10^{-2}(M/M_\odot)^{1/2}$ pc 
in a $\om =0.3, \ola =0.7$ and $h=0.65$ cosmology.
Also, the Einstein radius projected on the source plane should be larger
than the size of the source. A Type Ia SN at maximum luminosity has a size of 
$\sim 10^{15}$ cm, implying a lower mass limit of $\sim 10^{-4} M_{\odot}$ for the 
case described above.  

The effects of lensing by compact objects are different from those of
lensing by halos consisting of smoothly distributed dark matter,
such as in the singular isothermal sphere or Navarro-Frenk-White 
density profile (NFW; \cite{nfw}),
the main difference being the tail of large magnifications caused by
small impact-parameter lines of sight near the compact objects. However,
N-body simulations predict, besides the overall cuspy profiles of
ordinary Galaxy-sized dark matter halos, also a large number of small 
subhalos on all length scales which can been resolved \cite{nfw,moore}.
The number density of the smaller objects, of mass $M$,
 follows approximately the law
$dN/dM\propto M^{-2}$ as predicted by Press-Schechter theory. 
Thus one may expect a multitude of subhalos in each galaxy or cluster
halo. In addition,
N-body simulations show the less massive halos to be denser (mainly due
to them being formed early when the background density was higher).
Thus, it is appropriate to address the question whether this type of 
small-scale structure, and in particular the dense central regions of them, 
may give rise to lensing effects similar to
the ones caused by truly compact objects.

To put a bound of these possible effects, we use the results of
the most accurate numerical simulations to date \cite{moore}.  
The density profiles within
dark matter clumps obtained in the simulations can be fit by the Moore profile
\beq
\rho_M(r)={\rho'_M\over \left(r\over a\right)^{1.5}
\left(1+\left(r\over a\right)^{1.5}\right)}
\eeq
Here $\rho'_M$ and $a$ are not independent parameters, but related
by the concentration parameter $c_M=R_{200}/a$, which depends on
mass roughly as
\beq
c_M\sim 10.6M_{12}^{-0.084}.
\eeq
(Here $M_{12}$ is the virial mass in units of $10^{12}$ $M_\odot$; $R_{200}$
is the virial radius where the average overdensity is 200 times the background
density. Similar relations appear in the NFW simulations.)

Using these relations we can derive the mass $M_R$ (also in units
of $10^{12}$ $M_\odot$) within distance $R_{pc}$ parsecs 
from the center of the 
halo,
\beq
M_R\sim 10^{-7}M_{12}^{0.4}R^{1.5}_{pc}
\eeq
Comparing this with the Einstein radius for the same mass for a lensing
event of typical distance $D_{Gpc}$ Gigaparsecs,
\beq
R_E\sim 10^4 M_R^{0.5}D^{0.5}_{Gpc}\ {\rm pc},
\eeq
we find that $M_R$ is within its own Einstein radius
if
\beq
M_R\lesssim 10^{-4}M^{1.6}_{12}D_{Gpc}^3.\label{eq:lensmoore}
\eeq
The requirement that the lensing  mass is
greater than $10^{-4}M_\odot$ means
that dark matter clumps of the Moore type with mass greater than around
$10^4M_\odot$ ($M_{12} > 10^{-8}$) will in principle contribute to 
compact lensing.
However, it is only the very central, dense core that can contribute
to the lensing, and we see from Eq.\,~(\ref{eq:lensmoore})
that only a small fraction $f$ of the mass in the clump
is within its own Einstein radius,
\beq
f={M_R\over M_{12}}\sim 10^{-4}M_{12}^{0.6} D^3_{Gpc}.
\eeq
Thus only clumps of Galaxy size contribute appreciably to the
lensing, and this part is included in our standard calculations.

Since this analysis was made for Moore-type halos, which are more 
concentrated than NFW halos, we conclude that compact objects detected
through lensing of supernovae cannot, according to current thinking
about structure formation, be caused by clumps of particle dark matter
formed through hierarchical clustering.

\section{Gravitational lensing of supernovae}\label{sec:sn}
\begin{figure}[t]
  \centerline{\hbox{\epsfig{figure=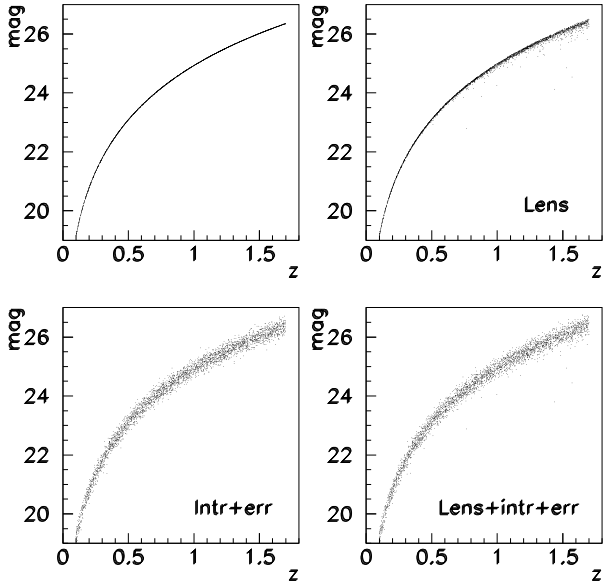,width=0.5\textwidth}}}
  \caption{A comparison of the dispersion due to lensing (upper right) and the 
    measurement error and intrinsic dispersion of Type Ia SNe (lower left).}
  \label{fig:dispersion} 
\end{figure}

The effect of gravitational lensing on Type Ia SN measurements is to cause a dispersion in
the Hubble diagram. In Fig.~\ref{fig:dispersion} we compare the dispersion due to gravitational 
lensing with the intrinsic dispersion and the typical measurement error for Type Ia SNe. 
In the upper left panel, we show an ideal Hubble diagram
with no dispersion and in the upper right panel we have added the dispersion due to 
lensing (lens) in a universe
with 20 \% compact objects and 80 \% smooth dark matter halos parametrized by the 
Navarro-Frenk-White formula \cite{nfw}. 
Comparing with the panel in the lower left where the intrinsic dispersion (intr) and 
measurement error (err) have been included, we see that the effects become comparable at a 
redshift of unity. In the lower right we
see the most realistic simulation with intrinsic dispersion, measurement error and 
lensing dispersion.

Of course, the additional dispersion caused by gravitational lensing will be a 
source of systematic error in the cosmological parameter determination with 
Type Ia supernovae. However, a possible virtue of lensing is that the 
distribution of luminosities might be used to obtain some information on the 
matter distribution in the Universe, e.g., to determine the fraction of compact
DM in our Universe.

\section{SuperNova Observation Calculator}\label{sec:snoc}

To perform realistic calculations we have developed a numerical simulation
package, SNOC (the SuperNova Observation Calculator). 
It can be used to
estimate various systematic effects such as dust extinction and gravitational lensing on
current SN measurements as well as the accuracy to which various parameters can be measured 
with future SN searches. 
In this paper, we use SNOC to obtain simulated samples of the intrinsic dispersion and
gravitational lensing effects of Type Ia SNe over a broad redshift range. 

The intrinsic dispersion and measurement error is represented by a Gaussian distribution 
with $\sigma_m =0.16$ mag. 
Gravitational lensing effects are calculated by tracing the light between the source and 
the observer by sending it through a series of spherical cells in which the dark
matter distribution can be specified. 
For more details on the method, originally proposed by Holz and Wald, 
see \cite{holz,bergstrom}.

We will model compact DM as point-masses
and smooth DM as the Navarro-Frenk-White density profile. The exact parameterization
of the smooth DM halo profile is not important for the results obtained in this paper
\cite{bergstrom}.
The results are also independent of the individual masses of the compact objects as well
as their large scale clustering properties \cite{holz,bergstrom}. As we have seen,
the eventual small-scale structure in the ``smooth'' component does
 not act as a compact component.

\section{Supernova/Acceleration Probe}\label{sec:snap}
To make realistic predictions of the statistics and quality of the supernova
sample, we use the projected discovery potential of 
the Supernova/Acceleration Probe (SNAP) project. This is a proposed 
satellite telescope capable of discovering
over 2000 Type Ia SNe per year in the redshift range $0.1<z<1.7$ \cite{snapprop}. 
The most anticipated use of the data is  to gain
further accuracy in the determination of, e.g., $\om$ and the curvature of the Universe
$\Omega_{\rm k}$, but also to give insight into the nature of the negative pressure energy 
component by constraining the equation of state \cite{huterer,goliath2}, or
the redshift dependence of the effective energy density \cite{max}. 
Here we show how the data also can be used to give information on the fraction
of compact objects in the dark matter component. 

\begin{figure}[t]
  \centerline{\hbox{\epsfig{figure=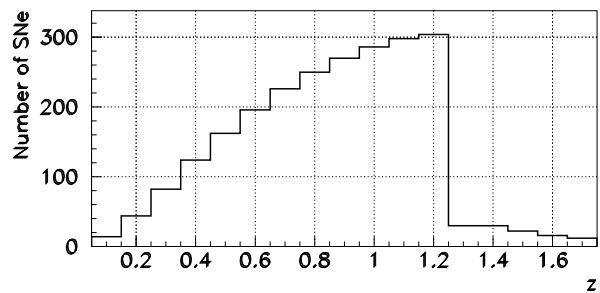,width=0.5\textwidth}}}
  \caption{Number of Type Ia SNe expected for various redshift bins in a one-year exposure 
    with the proposed SNAP satellite. The data are taken from Table 7.2 in the SNAP science 
    proposal \protect\cite{snapprop}.}\label{fig:snapdata} 
\end{figure}

More specifically, it is projected that in one year, SNAP will be able to discover, 
follow the light curves and obtain spectra for of the order of 2000
SNe. While the exact redshift distribution of the Type Ia supernovae to be followed
by SNAP might be changed upon studies of the optimal search strategies for the primary goals
of the project, we have used the Monte-Carlo generated sample
spectra of 2366 SNe distributed according to the SNAP proposal
\cite{snapprop}, see Fig.~\ref{fig:snapdata}, to make specific predictions. 

\section{Results}\label{sec:res}
\begin{figure}[t]
  \centerline{\hbox{\epsfig{figure=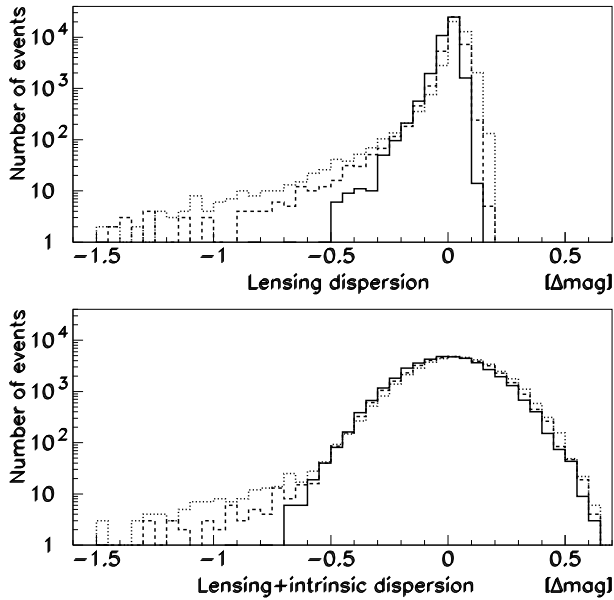,width=0.5\textwidth}}}
  \caption{Magnitude dispersion of reference samples for 0 \% (full line), 
20 \% (dashed line) and 40 \% (dotted line) compact objects. 
The bottom panel includes a Gaussian smearing, $\sigma_{\rm m}=0.16$ mag, due to 
intrinsic brightness differences between supernovae and from measurement error. 
The distributions show the projected 
scatter around the ideal Hubble diagram for Type Ia supernovae with a relative
redshift distribution as in Fig.~\ref{fig:snapdata}.}\label{fig:art1x2} 
\end{figure}

Using SNOC, we have created large data sets of synthetic 
SNe observations with a variable fraction
of compact objects ranging from 0 to 40 \%
using the following cosmological background parameter values: $\om =0.3, 
\ola =0.7$ and $h=0.65$. (For a discussion of how the halo distributions
were generated, see \cite{bergstrom}.)
These are used as reference samples.
In Fig.~\ref{fig:art1x2}, we have plotted the dispersion in the reference samples for 0 \% (full line), 
20 \% (dashed line) and 40 \% (dotted line) compact objects. 
In the upper panel, the lensing
dispersion is plotted, i.e., the dashed line basically corresponds to the scatter in the upper 
right panel in Fig.~\ref{fig:dispersion}. 
The zero value corresponds to the value one would obtain in a
homogeneous universe (upper left panel in Fig.~\ref{fig:dispersion}). 
Note that negative values correspond to magnified events, positive values
to demagnified events. 
It is clear, that as the fraction of compact DM grows, lensing effects 
becomes larger in the sense that we have a larger fraction of 
highly magnified events. In the 
lower panel we have added a Gaussian intrinsic dispersion and measurement error, 
$\sigma_{\rm m}=0.16$ mag, making the 
distributions look more 
similar (cf., lower right panel in Fig.~\ref{fig:dispersion}).
Although this smearing obviously decreases the significance of the
compact signal, it can be seen that the high-magnification tails still are
clearly visible.

\begin{figure}[t]
  \centerline{\hbox{\epsfig{figure=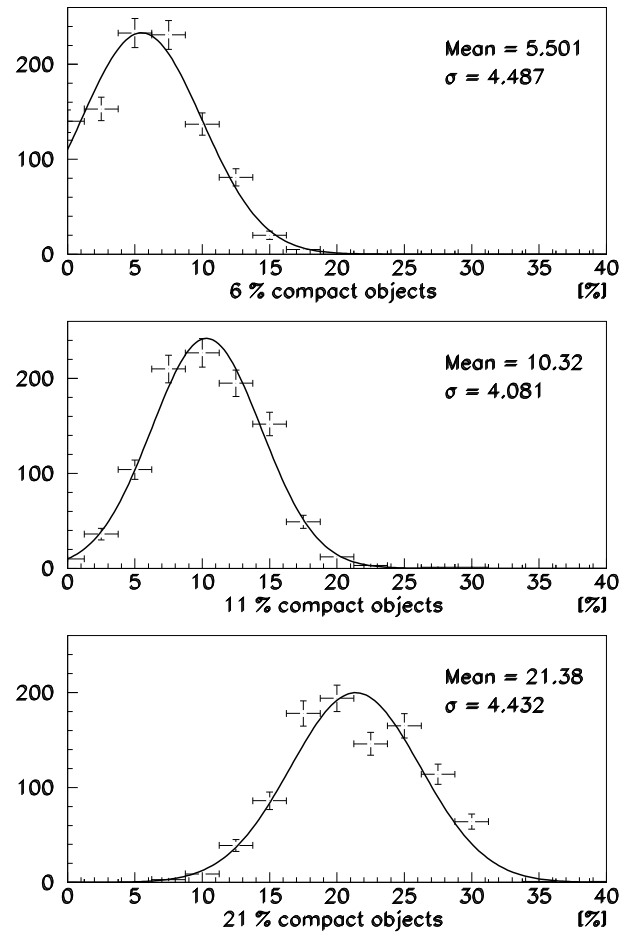,width=0.5\textwidth}}}
  \caption{The number of best-fit reference samples as a function of the 
    fraction of compact objects in the reference sample.}\label{fig:art1x3} 
\end{figure}

We have also created a large number of simulated one-year SNAP data sets 
(according to Fig.~\ref{fig:snapdata} above) with 6, 11 and 21
\% compact objects. These are our experimental samples. 
By comparing each generated experimental sample with our 
high-statistics reference samples using the Kolmogorov-Smirnov 
(KS) test, we obtain a confidence level for the hypothesis 
that the experimental sample is drawn from the same
distribution as the reference sample. 
For each fraction of compact objects in the experiments
(6, 11 and 21 \%), we repeat this procedure for 1000 
experimental realizations and
pick out the reference sample which gives the highest confidence level for 
each experiment.
Plotting the number of best-fit reference samples as a function of the 
fraction of compact objects 
in the reference sample, we can fit a Gaussian and thus estimate the 
true value and the dispersion.
In each case, we get a mean value 
within 1 \% of the true value and a one-sigma error less
than 5 \%, see Fig.~\ref{fig:art1x3}.

As one could expect, and is shown 
in Fig.~\ref{fig:dispersion}, lensing effects get larger at higher redshifts.
At low redshifts the dispersion is completely dominated by the 
intrinsic dispersion. Therefore, 
we have only used the data from SNe at $z>0.8$, a total of 1387 SNe, 
to obtain the result in Fig.~\ref{fig:art1x3}.\footnote{Of course, 
the data from SNe at $z<0.8$ can still be used to constrain the values of $\om$
and $\ola$. In fact, with one year of SNAP data, it is possible to determine $\om$ with a
statistical uncertainty of $\Delta\om\approx$ 0.02 \protect\cite{snapprop}.}

\begin{figure}[t]
  \centerline{\hbox{\epsfig{figure=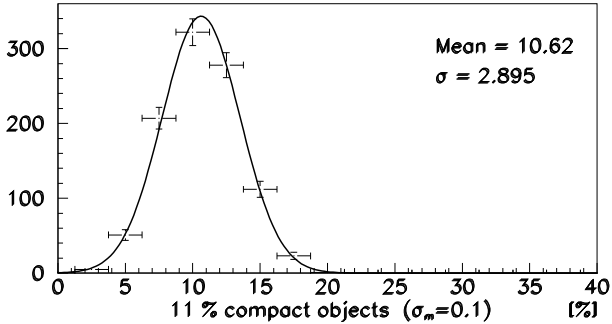,width=0.5\textwidth}}}
  \caption{The number of best-fit reference samples as a function of the 
    fraction of compact objects in the reference sample for an intrinsic dispersion and 
    measurement error of $\sigma_{\rm m}=0.1$ mag.}\label{fig:art1x1} 
\end{figure}

In this analysis, we have assumed that $\om$ and $\ola$ 
will be known to an accuracy
where the error in luminosity is negligible in comparison to the 
intrinsic and lensing 
dispersion of Type Ia SNe. This assumption is not unreasonable with future CMBR observations 
combined with other cosmological tests and the SNAP data itself, nor is it 
crucial in the sense that we are dealing with the dispersion 
of luminosities around the true mean value, not the mean value itself. 
In order to test the sensitivity of our results to changes in the cosmological parameter values,
we have performed a number of Monte-Carlo simulations using different sets of parameters in
our experimental and reference samples and found the error in the total energy density
in compact objects to be negligible\footnote{Note that the effect from lensing is
proportional to the total energy density in compact objects, not the fraction of compact
objects, see, e.g., \cite{SEF}. A higher $\om$ would therefore increase the  
sensitivity for smaller fractions.}.
Also, we have used a value of the intrinsic dispersion and measurement error of 
$\sigma_{\rm m}=0.16$ mag, a value which may be appreciably smaller in the future when the large 
data sample expected may allow, e.g., a more refined description and classification of supernovae.
Using simulated data sets with $\sigma_{\rm m}=0.1$ mag, the one-sigma error in the determination 
of the fraction of compact objects, using the same cosmology as above, 
becomes less than 3 \%, as depicted in Fig.~\ref{fig:art1x1}.

\section{Summary}\label{sec:disc}
The proposed SNAP satellite will be able to detect and obtain spectra for more than 2000
Typa Ia SNe per year. In this paper we have used simulated data sets obtained with the SNOC
simulation package to show how one-year SNAP data can be used to determine
the fraction of compact DM in our Universe to $\lesssim 5 \%$ accuracy, assuming the 
intrinsic dispersion and measurement error is $\sigma_{\rm m}=0.16$ mag. 
If the intrinsic dispersion and measurement error 
can be further reduced, e.g., from a improved understanding of 
Type Ia SN detonation mechanisms, the accuracy can be improved even further.  

\section*{Acknowledgements}
LB wishes to thank the Swedish Research Council for financial 
support, and the Institute for Advanced Study at Princeton for 
hospitality when part of this work was performed. 
AG is a Royal Swedish Academy 
Research Fellow supported by a 
grant from the Knut and Alice Wallenberg Foundation.

\end{document}